\documentstyle[preprint,psfig,epsfig,graphics,revtex]{aps}

\begin{document}
\draft
\preprint{}
\begin{title}
Correlation function of random heteropolymer solutions
\end{title}

\author{Zh. S. Gevorkian$^{1,2,*}$  and   Chin-Kun Hu$^{1,+}$}

\begin{instit}
$^1$Institute of Physics, Academia Sinica,
Nankang, Taipei 11529, Taiwan\\
$^2$Institute of Radiophysics and Electronics\\
and Yerevan Physics Institute, Alikhanian Br.2\\
375036 Yerevan Armenia
\end{instit}

\vskip 1.0 cm
\begin{abstract}
We study the density-density correlation 
function of the dense random heteropolymer  solutions.
We show that a phase transition is possible due to the 
heterogeneity of polymers. We also show that the critical behavior
 of the system is described by the $O(N)$ model at $N=0$.
\newline

PACS number(s): 61.25.Hq

\end{abstract}

\thispagestyle{empty}

\newpage

\section{Introduction}

\indent

Random heteropolymers are of great interests because of 
their wide occurence in nature, e.g. many biopolymers 
are random heteropolymers. It is well known that they have many unusual
properties \cite{Shakh,Gros,TKD}. For example, when the 
temperature  decreases they experience microphase separation 
and freezing phase transitions.
The protein folding is also an example of such a property. However
all of these properties refer mainly to a single heteropolymer.
The properties of polymer solutions have been well studied 
\cite{Edv,degen}. One of the well known
properties is the screening of  excluded volume interactions in
polymer solutions. It leads to an exponential decrease of the
density-density correlation function with  distance. In two dimensions,
 some peculiarities appear in the behavior of the correlation function. 
It turns out that the excluded volume effects lead to a phase transition in
a dense polymer solution \cite{Dup}. The ordered phase is described by the 
low temperature phase of  $N=0$ vector model \cite{Nie}. One of the
features of the ordered phase is that the correlation function increases
with  distance.

 Along with the single heteropolymers and polymer solutions, it is 
interesting to investigate also heteropolymer solutions. In this paper 
this problem is investigated for the first time. We will see below that
 a phase transition is possible due to the heterogeneity of polymer chains.

The paper is organized as follows. In Section II we derive an 
equation for the correlation function of a random heteropolymer solution.
 In Section III we develop a diagram technique
to treat with the heterogeneity random field. A critical heterogeneity
parameter is introduced in Section IV. In Section V we show that the
crossed diagrams for self-energy are smaller than the
non-crossed ones. Then we sum the non-crossed diagrams and improve
 the mean field approximation. In Section VI we present a field-theoretical
 approach and show that the problem is reduced to the $O(N)$ model at $N=0$.
 Section VII concludes our results. In the Appendix, we justify
the approximations done in Section II.

\section{Initial Relations}
  
 The correlation function of a polymer solution is determined as follows
\begin{equation}
G(\vec{r})=\frac{1}{c}[<c(\vec{r})c(0)>-c^2],
\label{In1}
\end{equation}
where
\begin{equation}
c(\vec{r})=\sum_{i}\delta(\vec{r}-\vec{R_i}),
\label{In2}
\end{equation}
is the microscopic concentration of monomers, $\vec{R_i}$ is the coordinate
 of the $i-th$ monomer, $<...>$ means the thermodynamic averaging and $c$
 is an average concentration of monomers. Note that Eq.(\ref{In2}) 
implies a summation over  all chains and monomers.
Now let us derive an equation for the correlation function (\ref{In1}).
It can be represented in the form \cite{Edv}
\begin{equation}
cG(\vec r)=A\int\prod d\vec R_i c(\vec r)c(0)exp\left[-\frac{U}{T}\right],
\label{In3}
\end{equation}
where $U$ is the conformational energy of polymers and $A$ is a normalization constant
\begin{equation}
A=\int\prod d\vec R_i exp\left[-\frac{U}{T}\right].
\label{In4}
\end{equation}
The conformational energy consists of elastic and excluded volume parts
\begin{equation}
\frac{U}{T}=\frac{\alpha}{2a^2}\sum_{bn}\left(\vec R_{b n}-\vec R_{b n-1}\right)^2
+\frac{1}{2}\sum_{ij}v_{ij}\delta(\vec R_i-\vec R_j).
\label{In5}
\end{equation}
Here $\alpha =3, 2$ for three and two dimensions, respectively, the index 
$b$ denotes a polymer chain and $n$ is the number of a monomer in a 
chain, $a$ is the average size of monomers, $i\equiv{b,n}$ and $v_{ij}$ 
are the random excluded volume constants. It is convenient in (\ref{In3}) 
to go to integration on the monomer concentration $c(\vec r)$ instead of $R_i$.
 For this purpose one has to express the conformational energy $U$ 
in terms of monomer concentration. It is well known \cite{Edv} that 
the elastic energy of a polymer solution, in the Gaussian
approximation, can be represented in the form
\begin{equation}
\frac{U_{el}}{T}=\frac{BV}{2}\sum_{\vec k}a^2k^2c_{\vec k}c_{-\vec k}
\label{In6}
\end{equation}
where  $B=1/12c,1/8c$ for three \cite{Edv} and two dimensions, 
respectively, $V$ is the volume
of the system and $c_{\vec k}$ is the Fourier transform of $c(\vec r)$
\begin{equation}
c_{\vec k}=\frac{1}{V}\int c(\vec r)e^{i\vec k\vec r}d\vec r.
\label{In7}
\end{equation}
Note that the approximation (\ref{In6}) is justified in the concentrated
solutions where the density fluctuations are small \cite{Edv}.
It is convenient to represent the random excluded volume constant in the form
 $v_{ij}=v+\delta v_{ij}$, where $v$ is the average excluded volume constant
 and $\delta v_{ij}$ is the fluctuating part with the average $<\delta v_{ij}>=0$. Using
Eqs.(\ref{In2}) and (\ref{In5}) for the average part of excluded volume energy, one finds
\begin{equation}
\frac{U_{exc}^{av}}{T}=\frac{v}{2}\int d\vec r c^2(\vec r).
\label{In8}
\end{equation}
It is shown in the Appendix that if $\delta v_{ij}$ are independent random 
variables with the variance $<\delta v_{ij}^2>=w^2$  
\cite{Roan} then the fluctuating part of excluded volume energy can be
 represented in the form
\begin{equation}
\frac{U_{exc}^1}{T}=\frac{1}{2}\int v_1(\vec r)c^2(\vec r)d\vec r,
\label{In9}
\end{equation}
where $v_1(\vec r)$ is a Gaussian distributed random function with $\delta$ correlations
\begin{equation}
<v_1(\vec r)v_1(\vec r^\prime)>=\frac{w^2}{c}\delta(\vec r-\vec r^\prime),\quad <v_1>=0.
\label{In10}
\end{equation}
Using Eqs.(\ref{In6}), (\ref{In8}) and (\ref{In9}) for the conformational energy
 of a random heteropolymer solution, finally we obtain
\begin{equation}
\frac{U}{T}=\frac{U_{el}+U_{exc}^{av}+U_{exc}^1}{T}
=\frac{1}{2}\int d\vec r c(\vec r)\left[-Ba^2\nabla^2+v+v_1(\vec r)\right]c(\vec r).
\label{In11}
\end{equation}
Now the correlation function (\ref{In1}) can be written in terms
 of a functional integral over $c(\vec r)$  \cite{Edv}
\begin{equation}
cG(\vec r)=A\int Dc(\vec r) c(\vec r)c(0)exp\left[-\frac{U[c(\vec r)]}{T}\right],
\label{In12}
\end{equation}
where
\begin{equation}
Dc(\vec r)=\prod_{k>0} dc_{\vec k}.
\label{In13}
\end{equation}
The symbol $\prod_{k>0}$ means an integration over the independent components 
$c_{\vec k}$. Because the relation $c_{\vec k}^*=c_{-\vec k}$ (this follows 
from the fact that $c(\vec r)$ is
a real function) not all the components of $c_{\vec k}$ are independent.
Calculating the Gaussian integral over $c(\vec r)$ in Eq.(\ref{In12}),
one finds
\begin{equation}
cG(\vec r)=\left[-Ba^2\nabla^2+v+v_1(\vec r)\right]^{-1}\delta(\vec r).
\label{In14}
\end{equation}
It is easy to obtain the equation for $G(\vec r)$ from Eq.(\ref{In14})
\begin{equation}
\left[\nabla^2-\xi^{-2}-\xi^{-2}\frac{v_1(\vec r)}{v}\right]G(\vec r)
=-\frac{\xi^{-2}}{vc}\delta(\vec r).
\label{In15}
\end{equation}
Here $\xi=(Ba^2/v)^{1/2}$ is the correlation length of the polymer solution. 
 
\section{Impurity Diagram Technique}

For bare correlation function with $v_1=0$, which corresponds to 
homopolymer solution, one has from (\ref{In15})
\begin{equation}
G_0(q)=\frac{\xi^{-2}}{vc(q^2+\xi^{-2})}.
\label{CF12}
\end{equation}
In the coordinate representation we have from Eq.(\ref{CF12})
\begin{equation}
G_0(\vec{r})=\frac{\xi^{-2}}{vc}\int \frac{d\vec{q}}{(2\pi)^d}
\frac{exp(i\vec{q}\vec{r})}{q^2+\xi^{-2}}.
\label{CF13}
\end{equation}
Using the method of steepest descent in the integral of
Eq. (\ref{CF13}) over the momentum \cite{ZJ}, we can find
the asymtotic behavior of correlation function on large distances
 
\begin{equation}
G_0(\vec r)\sim \frac{\xi^{-1}}{2vc}(\frac{\xi^{-1}}{2\pi r})^{\frac
{d-1}{2}}e^{-\frac{r}{\xi}}.
\label{CF14}
\end{equation}
It follows from Eq.(\ref{CF14}) that the correlation function at large
distances decreases exponentially  with the correlation length $\xi$.

Now we  construct a perturbation theory with the
heterogeneity parameter $v_1$.
It is convenient to deal with the dimensionless in $d=2$ correlation
function $G(\vec r) \equiv G(\vec r)vc\xi^2$ that satisfies the
equation
 \begin{equation}
\left[\nabla^2-\xi^{-2}-\xi^{-2}\frac{v_1(\vec{r})}{v}\right]G(\vec{r})=-
\delta(\vec{r}).
\label{Di18}
\end{equation}
In order to construct a perturbation theory in the heterogeneity
parameter, we use  the impurity diagram technique \cite{AGD}. It is easy
to obtain the following expansion from Eq.(\ref{Di18})
\begin{equation}
\parbox[c][3cm][c]{12cm}{\includegraphics[scale=0.65]{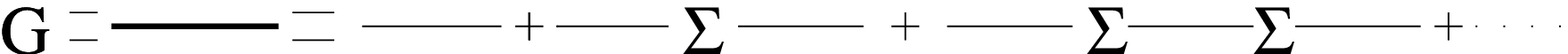}},
\label{Di19}
\end{equation}
where the self-energy $\Sigma$ is determined by the following
irreducible diagrams
\begin{equation}
\parbox[c][3.5cm][c]{13cm}{\includegraphics[scale=0.65]{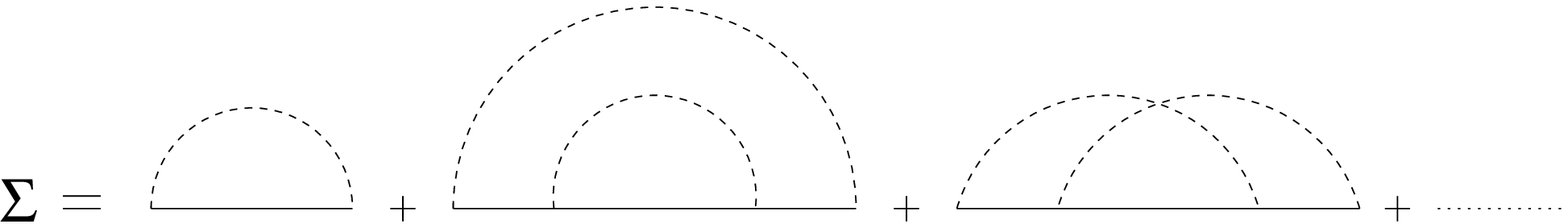}}.
\label{Di20}
\end{equation}  
The heavy line in Eq.(\ref{Di19}) denotes the averaged correlation function,
 the thin one denotes the bare correlation function of
Eq. (\ref{CF12}) and the dashed one  is the Fourier transform of the
 heterogeneity random field correlation function of Eq.(\ref{In15}),
 $\xi^{-4}w^2/(cv^2)$. Summing the diagrams in Eq.(\ref{Di19}), 
we have the following Dyson equation
\begin{equation}
\parbox[c][3cm][c]{12cm}{\includegraphics[scale=0.65]{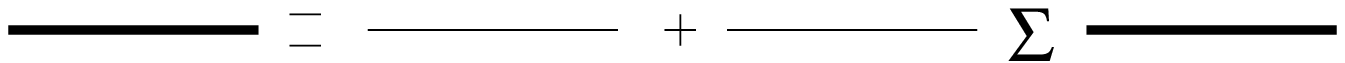}}.
\label{Di21}
\end{equation}

It has a solution 
\begin{equation}
G(q)=\frac{1}{q^2+\xi^{-2}-\Sigma(q)},
\label{Di22}
\end{equation}
where the self-energy is determined by  diagrams
of Eq. (\ref{Di20}). In the leading order, we get for the self-energy 
\begin{eqnarray}
\Sigma_0(q)=\frac{\xi^{-4}w^2}{cv^2}\int
\frac{d\vec{k}}{(2\pi)^d}G_0(\vec{q}-
\vec{k})&=&
\frac{\xi^{-4}w^2}{4\pi cv^2}\ln(1+\frac{\xi^2}{v})~~{\rm for} \quad d=2,~{\rm and}
\nonumber \\
&=& \frac{\xi^{-4}w^2}{2\pi^2 c v^2}(\frac{1}{v^{1/3}}-
\frac{\pi}{2\xi})~~{\rm for} \quad d=3.           
\label{Di23} 
\end{eqnarray}

The divergence of the integral in the upper limit in Eq.(\ref{Di23})
is caused by the $\delta$ feature of the excluded volume interaction of
Eq.(\ref{In5}). It is obvious that it should be smoothed on the scales
less than $v^{1/d}$. Therefore we cut the upper limit of the integral
in Eq.(\ref{Di23}) by this value.

It follows from Eqs.(\ref{Di22}) and (\ref{Di23}), that the ``mass'' term 
in the denominator of correlation function of Eq. (\ref{Di22}) is renormalized as
\begin{equation}
 \xi_R^{-2}= \xi^{-2}-\Sigma.
\label{Di24}
\end{equation}
It follows from Eqs.(\ref{Di22})-(\ref{Di24}), that the second term in
Eq.( \ref{Di24}) at some critical value of heterogeneity 
parameter $w^2/v^2$ can become equal to the first one. In this case
the correlation length diverges and the exponential decay of 
the correlation function of Eq. (\ref{CF14}) at large distances will be 
substituted by a power law. This is a second order phase transition 
caused by the heterogeneity of polymers.

\section{Critical Heterogeneity Parameter}

 It is easy to obtain the critical value of the heterogeneity parameter
$g=w^2/v^2$ from the condition of divergence of the renormalized correlation 
length. Using Eqs. (\ref{Di23})-(\ref{Di24}), one has
\begin{eqnarray}
g_c&=& \frac{4\pi c\xi^2}{\ln(1+\xi^2/v)}~~ {\rm for} \quad d=2,~{\rm and} \nonumber\\
   &=& \frac{2\pi^2c\xi^2}{1/v^{1/3}-\pi/2\xi}~~{\rm for} \quad d=3.
\label{Hp25}
\end{eqnarray}

Taking into account that almost always $\xi \gg v^{1/3}$, we can simplify
 the expression of Eq. (\ref{Hp25}) by
\begin{eqnarray}
g_c &=& \frac{\pi a^2}{4v\ln\frac{a^2}{v\sqrt{8\Phi}}}~~{\rm for} 
\quad d=2,~{\rm and} \nonumber\\
    &=& \frac{\pi^2}{6}(\frac{a}{v^{1/3}})^2~~{\rm for} \quad d=3.
\label{Hp26}
\end{eqnarray}

where $\Phi=ca^2$ is the fraction of monomers in the solution. It follows
from Eq.(\ref{Hp26}) that $g_c$ does not depend on monomer concentration
(only logarithmically in two dimensions). It is mainly determined by the
stiffness parameter $v/a^3$ ($v/a^2$ in two dimensions) of polymer chains.
It is also obvious from Eqs.(\ref{Hp25})-(\ref{Hp26}) that the phase
transition in heterogeneity parameter is more appropriately realized
in a system of flexible chains $v/a^3\le 1$ rather than in stiff 
$v/a^3\ll 1$ chains solution because the limit $g\to g_c$ is easier
 reached in the first case.
One can obtain from Eqs.(\ref{Di22})-(\ref{Di24}) that near the 
critical point the renormalized correlation length diverges as
\begin{equation}
\xi_R \sim \xi (\frac{g_c-g}{g_c})^{-1/2}
\label{Hp27}
\end{equation}
So, for the critical exponent of the correlation length in the phase
$g<g_c$ we obtain the mean field result. This means that tree approximation 
that we used in Eqs. (\ref{Di19}), (\ref{Di22}) and (\ref{Di23}) for correlation
function corresponds to the mean field theory. At the critical point $g=g_c$
one has power law behavior for correlation function on large distances
instead of the exponential decreasing in $g<g_c$ case. In the next 
section we try to go beyond the mean field approximation. 

\section{Beyond the Mean Field Approximation} 

Let us consider the second order diagrams in the diagrammatic expansion
of the self-energy in Eq.(\ref{Di20}). For the contribution of the second
diagram in (\ref{Di20}), one has
\begin{equation}
\Sigma_2(q)=(\frac{\xi^{-4}w^2}{cv^2})^2\int\frac{d\vec{k_1}d\vec{k_2}}
{(2\pi)^{2d}}G^2(\vec{k_1}-\vec{q})G(\vec{k_2}+\vec{k_1}-\vec{q}).
\label{BM28}
\end{equation}
We take the correlation functions in (\ref{BM28}) in the tree approximation
\begin{equation}
G(k)=\frac{1}{k^2+\xi_R^2}
\label{BM29}
\end{equation}
where the renormalized correlation length is determined by Eq.(\ref{Di24})
and $\xi_R\to\infty$ at the critical point $g=g_c$. Substituting Eq. (\ref{BM29})
 into Eq. (\ref{BM28}) and integrating over the momentums,
we have
\begin{eqnarray}
\Sigma_2 &=&
(\frac{\xi^{-4}w^2}{cv^2})^2\frac{\xi_R^2}{16\pi^2}
\ln\frac{\xi_R^2}{v}~~{\rm for} \quad d=2,~{\rm and} \nonumber\\
&=& (\frac{\xi^{-4}w^2}{cv^2})^2\frac{\xi_R}{8\pi^3v^{1/3}}~~{\rm for} \quad d=3.
\label{BM30}
\end{eqnarray}

When obtaining Eq.(\ref{BM30}) we cut the upper limit of momentum integrals in
 Eq.(\ref{BM29}) at $1/v^{1/d}$ and believe that 
$\xi_R\gg v^{1/d}$ which is always correct in the critical region $g\to g_c$.
Now consider the contribution of the third diagram in Eq.(\ref{Di20}).
It can be represented in the form
\begin{equation}
\Sigma_3(q)=(\frac{\xi^{-4}w^2}{cv^2})^2\int\frac{d\vec{k_1}d\vec{k_2}}
{(2\pi)^{2d}}G(k_1)G(\vec{k_1}+\vec{k_2})G(\vec{k_2}-\vec{q}).
\label{BM31}
\end{equation}
Substituting Eq.(\ref{BM29}) into Eq.(\ref{BM30}) and going to the
dimensionless variables of integration, one has
\begin{equation}
\Sigma_3(q)=(\frac{\xi^{-4}w^2}{cv^2})^2\xi_R^{2d-6}\int\frac{d\vec{k_1}
d\vec{k_2}}{(2\pi)^{2d}}\frac{1}{(k_1^2+1)(k_2^2+1)}\frac{1}{(\vec{k_1}+
\vec{k_2}+\vec{q})^2+1}.
\label{BM32}
\end{equation}
We are mainly interested in the large distance behavior of the correlation
function. Our analysis of the integral in Eq.(\ref{BM32}) at the limit $q\to
0$ shows that
\begin{eqnarray}
\Sigma_3(0)&\sim& (\frac{\xi^{-4}w^2}{cv^2})^2
\frac{\xi_R^2}{16\pi^2} ~~{\rm for} \quad d=2,~{\rm and}\nonumber \\
&\sim& (\frac{\xi^{-4}w^2}{cv^2})^2\frac{1}{8\pi^3}\ln\frac{\xi_R}{v^{1/3}}
~~{\rm for} \quad d=3.
\label{BM33}
\end{eqnarray}

Comparing Eq.(\ref{BM33})with the Eq.(\ref{BM30}) one can see that in the
 critical region $\xi_R\gg v^{1/d}$ the crossed diagrams ,because the 
additional integration over angles, (see also \cite{AGD}) have an additional
 smallness on parameter $v^{1/d}/\xi_R\ll 1$. So, in the critical region 
the non-crossed diagrams are the dominant ones and we can sum them
\begin{equation}
\parbox[c][3.5cm][c]{14cm}{\includegraphics[scale=0.9]{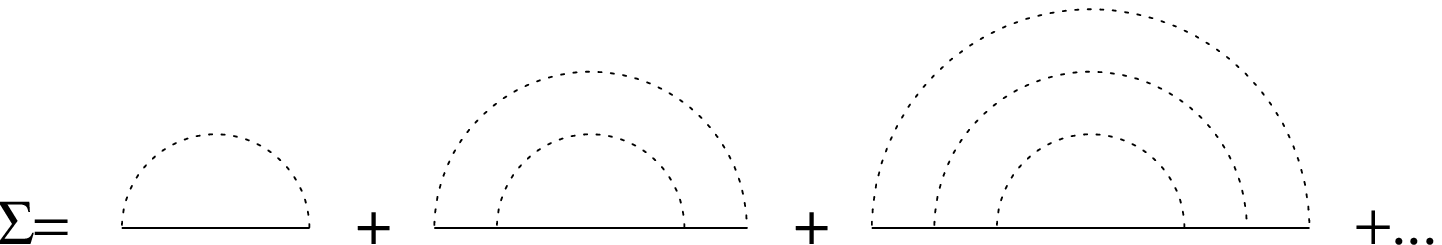}}.
\label{BMg}
\end{equation}
The sum of the diagrams can be represented in the form
\begin{eqnarray}
\Sigma_R=\frac{\Sigma_0}{1-\frac{\xi^{-4}w^2}{cv^2}\int\frac{d\vec{k}}
{(2\pi)^d}G^2(k)}&=& \frac{\Sigma_0}{1-\frac{\xi^{-4}w^2}{cv^2}
\frac{\xi_R^2}{4\pi}}~~{\rm for} \quad d=2,~{\rm and} \nonumber\\
&=& \frac{\Sigma_0}{1-\frac{\xi^{-4}w^2}{cv^2}\frac{\xi_R}{4\pi}}
~~{\rm for} \quad d=3.
\label{BM34}
\end{eqnarray}
The expressions are correct provided that 
$\xi^{-4}w^2\xi_R/4\pi cv^2< 1$. At the critical point $\xi_R\to \infty$.
Therefore, even this improved approximation which corresponds  the one-loop 
correction in field theoritical calculation still breaks down. To investigate 
the critical region $g\to g_c$ more carefully and to describe the ordered phase 
$g>g_c$, we use a field theoretical approach in the next section.

\section{Field Theoretical Approach}

It follows from Eqs.(\ref{Di18})-(\ref{Di24}) and (\ref{Hp26}) that the
renormalized correlation function satisfies the equation
\begin{equation}
\left[-\nabla^2+m^2+h(\vec{r})\right]G(\vec{r})=\delta(\vec{r}),
\label{FT35}
\end{equation}
where $m^2=\xi^{-2}(g_c-g)/g_c$ is  the renormalized ``mass'' term
and $h(\vec{r})=\xi^{-2}v_1(\vec{r})/v$ is the heterogeneity
random field. To carry out the averaging over the
realizations of random field, we use a functional integral to represent
 the correlation function:
\begin{equation}
G(\vec{r})=\frac{-i\int D\varphi(\vec{r})
\varphi(\vec{r})\varphi(o) \exp{S_0}}{\int
D\varphi(\vec{r})\exp{S_0}},
\label{FT36}
\end{equation}
where
\begin{equation}
S_0=-\frac{i}{2}\int d\vec{r}\varphi(\vec{r})\left[-\nabla^2+m^2+
h(\vec{r})\right]\varphi(\vec{r}).
\label{FT37}
\end{equation}
Here $\varphi(\vec{r})$ is a real field. Introducing the replica
fields and using Eq. (\ref{FT37}), one can represent the correlation 
function in the form
\begin{equation}
G(\vec{r})=\frac{-i\int\prod
D\varphi_\alpha(\vec{r})\varphi_1(\vec{r}) \varphi_1(0)\exp
S_N}{\int\prod D\varphi_\alpha(\vec{r})\exp S_N},
\label{FT38}
\end{equation}
where
\begin{equation}
S_N=-\frac{i}{2}\int
d\vec{r}\varphi_\alpha(\vec{r})\left[-\nabla^2+
m^2+h(\vec{r})\right]\varphi_\alpha(\vec{r}).
\label{FT39}
\end{equation}
In Eq.(\ref{FT39}), a summation over the repeated indexes is implied. 
Carrying out a Gaussian integration over $\varphi_\alpha$ in denominator
 of Eq.(\ref{FT38}), one has
\begin{equation}
\int\prod D\varphi_\alpha(\vec{r})\exp S_N=\left(\det
\left[\frac{-\nabla^2+ m^2+h(\vec{r})}{-2\pi
i}\right]\right)^{-N/2}.
\label{FT40}
\end{equation}
Finally, taking into account Eqs.(\ref{FT38}) and (\ref{FT40}) we
obtain the following representation for the correlation function
\begin{equation}
G(\vec{r})=\lim_{N\to 0}-i\int\prod
D\varphi_\alpha(\vec{r})\varphi_1(\vec{r}) \varphi_1(0)\exp S_N.
\label{FT41}
\end{equation}

Now using Eq.(\ref{FT41}), one can carry out the Gaussian averaging over
the random field $h(\vec{r})$
\begin{equation}
G(\vec{r})=\lim_{N\to 0}-i\int \prod
D\varphi_\alpha(\vec{r})\varphi_1(\vec{r})\varphi_1(0)\exp S,
\label{FT42}
\end{equation}
where
\begin{equation}
S=-\frac{i}{2}\int
d\vec{r}\varphi_\alpha(\vec{r})\left[-\nabla^2+m^2\right]
\varphi_\alpha(\vec{r})-\frac{H}{8}\int
d\vec{r}\left[\varphi^2_\alpha(\vec{r}) \right]^2.
\label{FT44}
\end{equation}
Here $H=\xi^{-4}w^2/(cv^2)$. So, we obtain a non-trivial
field-theoretical model. Expanding $\exp S$ in heterogeneity
parameter $H$ one can reproduce the impurity diagrams in Eq.
(\ref{Di20}).

So we map the concentrated random heteropolymer solution problem onto
the $O(N)$ vector model at $N=0$. This model is well studied 
\cite{ZJ} and we can use it's results. Remind that the problem
 of self-avoiding walk of a single polymer chain is also mapped to this
model \cite{deG}. However there is an important difference between these
 two problems. In our case the ordered phase $g>g_c$ (see below) also
exists in contrary  the former one where the criticality is
reached in the limit of long chains $L\to\infty$.

Thus, we have the following behavior for the correlation function in 
the phase $g<g_c$. When $g\ll g$ correlation function  at large distances 
exponentially decreases with some correlation length. At $g\to g_c$ 
correlation length diverges $\xi\sim (g_c-g)^{-\nu}$. The value of the
 exponent $\nu$ is well known from the $\epsilon-expansion$ and from the
 numerical calculations $\nu\approx 3/5,3/4$, for three and two dimensions,
 respectively. At the critical point the correlation function has a
power law behavior $G(r)\sim 1/r^{d-2+\eta}$. The exponent $\eta$ is
also well known for $O(N)$ model. This power law behavior is preserved 
also in the ordered phase in three dimenions. In two dimensions there are
 some peculiarities in the ordered phase. 
 It follows from the preceding consideration that the phase transition
is caused  by the ``massless'' fluctuations. In the vicinity of the critical
point we can simplify the action of Eq.(\ref{FT44}) by separating 
the ``massless'' fluctuations.
 For this reason we find saddle-point trajectories of the action
of Eq. (\ref{FT44}). Differentiating (\ref{FT44}), we obtain
\begin{equation}
i\nabla^2\varphi_\beta(\vec{r})-im^2\varphi_\beta(\vec{r})-\frac{H}{2}
\varphi_\beta(\vec{r})\sum_{\gamma}\varphi_\gamma ^2(\vec{r})=0. 
\label{FT45}
\end{equation}
We are looking for the homogeneous solution of Eq. (\ref{FT45}),
$\varphi_\beta(\vec{r})=const$
\begin{equation}
-im^2\varphi_\beta-\frac{H}{2}\varphi_\beta
\sum_{\gamma}\varphi_{\gamma}^2=0.
\label{FT46}
\end{equation}
This equation has two solutions
\begin{equation}
\varphi_\beta=0, \qquad
\sum_{\gamma}\varphi_{\gamma}^2=-\frac{2im^2}{H}, \qquad
\beta,\gamma=1,....N. 
\label{FT47}
\end{equation}

It can be seen from Eq.(\ref{FT47}) that the second type of saddle points 
lies in the complex plane. In the stationary phase method it is well known
that in such cases one must transform the integration contour so
that it passes through the saddle points (Fig. 1).
We have in mind that the integration contour is transformed
locally at each point $\vec{r}$. Along the new contour $\varphi$ has
the form $\varphi=\varphi^R (1-i)$, where $\varphi^R$ is real.
 Suppose that the fluctuating fields locally satisfy the extremum 
condition of Eq. (\ref{FT47}). Making a change
of variables $\varphi_\alpha(\vec{r})=\sqrt{m^2/H}\varphi_\alpha^R(1-i)$,
we  obtain the following expression for the effective
action describing the ``massless'' fluctuations
\begin{equation}
S_{eff}=-\frac{m^2}{H}\int d\vec{r}(\vec{\nabla}\varphi_\alpha)^2,
\label{FT48}
\end{equation}
where
\begin{equation}
\sum_\alpha \varphi_\alpha^2(\vec{r})=1.
\label{FT49}
\end{equation}
Here, for brevity, we omit the index $R$ and neglect the constant terms 
in action that do not affect the critical behavior. So, the critical
 behavior of random heteropolymer solution is described by the non-linear
 $\sigma$ model at $N=0$. The coupling constant of the model of 
Eq. (\ref{FT48}) is $m^2/H=c\xi^2(g_c-g)/(gg_c)$. The region $g<g_c$ 
 corresponds to the high temperature phase of $N=0$ vector model \cite{Sal}.
 Note that our derivation of $\sigma$ model from the $(\varphi^2)^2$ 
model is correct in the high temperature phase $g<g_c$. It is well
 known \cite{ZJ} that at the critical point the 
correlation functions of the non-linear $\sigma$-model
and $(\varphi^2)^2$ theory are identical. However $(\varphi^2)^2$ model
 is renormalizable at $d\le 4$ but $\sigma$ model is renormalizable 
only at $d\le 2$. So, in the critical domain for $d=2$ one can use the non-linear
$\sigma$-model instead of the $(\varphi^2)^2$-model.
The ordered phase $g>g_c$ will be described by the low-temperature phase
of the non-linear $\sigma$-model at $N=0$. However the coupling constant 
in this case differs from that of Eq.(\ref{FT49}). It will be 
proportional to the expectation value of the order parameter which is
 non-zero in the ordered phase \cite{ZJ}.
Note that this transition to the low-temperature phase of $N=0$ vector
 model in two-dimension is caused by the heterogeneity of chains and
 not by dense polymer self-avoiding walks \cite{Dup}. 
The low-temperature phase of $N=0$ vector model
has many unusual properties. One of them is  the increasing
of  spin-spin correlation function with  distance \cite{Nie}. In
our case this means that in the phase $g>g_c$ the correlation function
 will increase with  distance. The corresponding power index is $3/8$
\cite{Nie} . In the dense polymer case such a behavior is explained by 
the repelling of the extremities of  chains. To understand this
result for the heteropolymer solution, consider the case when each
chain consists of only two kinds of monomers (see Appendix). In
this case there will be three excluded volume constants $v_{aa}$,
$v_{ab}$ and $v_{bb}$. Suppose that the excluded volume constants
between monomers largely differ from each other so that $v_{ab}$, $v_{bb}
\gg v_{aa}$ (see also Appendix). In order to minimize  the
conformotional energy, type $a$ monomer  prefers contact with
type $a$ monomers. This means that the compensation of excluded volume
interactions leading to the screening effect in polymer solutions
will be violated in random heteropolymer solution. Moreover, the strong
repulsion between different kinds of monomers will lead to the
microphase separation like the analogous effect in single random
heteropolymers (see\cite{Shakh} and \cite{Gros} and references
therein). Although  these papers mainly dealt with the compact conformations 
of single random heteropolymers  (which is the case $v<0$, in our case
 always $v>0$), nevertheless we think that these two phenomena are 
closely related. The microphase  separation is the main reason of the
 rise of anti-correlations in heteropolymer solutions in two dimensions
 and power law decreasing of correlation function in three dimensions. 

\section{Conclusion}

We have considered the effect of random heterogeneity on the
density-density correlation function of dense polymer solutions. 
It turns out that a phase transition is possible due
to the heterogeneity of the polymer chains. The heterogeneity
parameter is determined by the ratio of variance and average
values of excluded volume constant. The critical behavior is described
by the O(N) vector model at $N=0$. The anti- correlation
behavior of correlation function in two dimensions and the power law
behavior in three dimensions in the ordered phase are
associated with the microphase separation.

This work was supported in part by the National Science Council of
the Republic of China (Taiwan) under the grant no. NSC
89-2112-M001-005.

\vskip 1.0 cm
\centerline{\bf Appendix: Random Excluded Volume Constant}
\vskip 1.0 cm
Using (\ref{In2}), one can represent the fluctuating part of excluded
volume energy in the form
\begin{equation}
\frac{U_{exc}^1}{T}=\frac{1}{2}\int d\vec r \sum_{ij} \delta v_{ij}
\delta(\vec r-\vec R_i)\delta(\vec r-\vec R_j).
\label{A1}
\end{equation}
Suppose that $\delta v_{ij}$ are independent random variables with
variance $w$. Then for the large number of monomers because of the 
central limit theorem, one has
\begin{equation}
\sum_j\delta v_{ij}\delta(\vec r-\vec R_j)\equiv v_i(\vec r)c(\vec r).
\label{A2}
\end{equation}
where $v_i$ are random Gaussian variables with  zero average
and variance $w$. To prove Eq. (\ref{A2}), we calculate the averages and variances
of left and right hand sides of Eq. (\ref{A2}). It is evident that $<v_i>=0$
because $<\delta v_{ij}>=0$. Now calculate the variance of the 
left hand side of Eq. (\ref{A2}),
\begin{equation}
\sum_{jk}<\delta v_{ij}\delta(\vec r-\vec R_j)\delta v_{ik}
\delta(\vec r-\vec R_k)>=\sum_{jk}<\delta v_{ij}\delta v_{ik}>
\delta(\vec r-\vec R_i)\delta(\vec r-\vec R_k).
\label{A3}
\end{equation}
Because $\delta v_{ij}$ are independent random variables, one has
\begin{equation}
<\delta v_{ij}\delta v_{ik}>=\delta_{jk}w^2.
\label{A4}
\end{equation}
Substituting Eq.(\ref{A4}) into Eq.(\ref{A3}), we have
\begin{equation}
\sum_{jk}<\delta v_{ij}\delta v_{ik}>\delta(\vec r-\vec R_j)
\delta(\vec r-\vec R_k)=w^2c^2(\vec r).
\label{A5}
\end{equation}
It is evident that the variance of right hand side of (\ref{A2})
 is also $w^2c^2(\vec r)$.
Substituting Eq.(\ref{A2}) into Eq.(\ref{A1}) for the fluctuating
 part of excluded volume energy, we have
\begin{equation}
\frac{U_{exc}^1}{T}=\frac{1}{2}\int d\vec r c(\vec r)\sum v_i(\vec R_i)
\delta(\vec r-\vec R_i).
\label{A6}
\end{equation}
Let us prove that
\begin{equation}
\sum_{i}v_i\delta(\vec r-\vec R_i)\equiv v_1(\vec r)c(\vec r),
\label{A7}
\end{equation}
where $v_1(\vec r)$ is a Gaussian distributed random function with
$\delta$-correlations. It is evident that $<\sum_iv_i\delta(\vec r-\vec R)>=0$.
 Now let us calculate the correlator
\begin{equation}
K(\vec r,\vec r^\prime)=\sum_{ij}<v_iv_j>\delta(\vec r-\vec R_j)
\delta(\vec r^\prime-\vec R_j).
\label{A8}
\end{equation}
The $v_i$ are independent Gaussian variables, therefore $<v_iv_j>
=\delta_{ij}w^2$. Substituting this into the Eq.(\ref{A8}), we have
\begin{equation}
K(\vec r,\vec r^\prime)=w^2\sum_{ij}\delta_{ij}\delta(\vec r-\vec R_i)
\delta(\vec r^\prime-\vec R_j).
\label{A9}
\end{equation}
For the dense solution of heteropolymers one can go to the continious limit
 $\delta_{ij}\to \delta(\vec R_i-\vec R_j)/c$,
\begin{equation}
K(\vec r,\vec r^\prime)=\frac{w^2}{c}\sum_{ij}\delta(\vec R_i-\vec R_j)
\delta(\vec r-\vec R_i)\delta(\vec r^\prime-\vec R_j).
\label{A10}
\end{equation}
Using the properties of $\delta$ functions, one has
\begin{equation}
K(\vec r,\vec r^\prime)=\frac{w^2}{c}\delta(\vec r-\vec r^\prime)c(\vec r)c(\vec r^\prime).
\label{A11}
\end{equation}
So the function $\sum v_i\delta(\vec r-\vec R_i)$ is equivalent to the function
$v_1(\vec r)c(\vec r)$. Finally, for the random part of excluded volume energy, we have
\begin{equation}
\frac{U_{exc}^1}{T}=\frac{1}{2}\int d \vec r v_1(\vec r)c^2(\vec r)
\label{A12}
\end{equation}
which coincides with Eq. (\ref{In9}).

 Now consider a special case
when each chain contains only two types of monomers.
 Suppose that each heteropolymer consists of $a$ and $b$ type
monomers with $1-f$ and $f$ fractions, respectively. Consider the
case  when $f\ll 1$ and $v_a\ll v_b$. Here $v_a\equiv v_{aa}$ and
$v_b\equiv v_{ab}$. As the effect of $v_{bb}$ will be proportional to
$f^2 \ll 1$ , we can ignore it. In such conditions the
average excluded volume constant can be estimated as
\begin{equation}
v=(1-f)v_a+fv_b.
\label{AA1}
\end{equation}
The variance (dispersity) of the excluded volume constant is estimated
as
\begin{equation}
w^2=f(1-f)(v_b-v_a)^2.
\label{AA2}
\end{equation}
The heterogeneity parameter for this type of randomness will be
\begin{equation}
g=\frac{w^2}{v^2}=\frac{f(1-f)(v_b-v_a)^2}{\left[(1-f)v_a+fv_b\right]^2}.
\label{AA3}
\end{equation}
Assume that $fv_b\ll (1-f)v_a$ then we get from Eqs. (\ref{AA1}) and
(\ref{AA3})
\begin{equation}
v\approx v_a, \qquad g\sim f\frac{v_b^2}{v_a^2}.
\label{AA4}
\end{equation}
Although $f v_b/v_a \ll 1$, the heterogeneity parameter can be large
\begin{equation}
g\sim \frac{f v_b}{v_a}\frac{v_b}{v_a}.
\label{AA5}
\end{equation}
because of $v_b \gg v_a$. As can be seen from Eq.(\ref{AA4}), in this
particular case the average excluded volume constant is determined
mainly by type $a$ monomers and the dispersity is determined by
type $b$ monomers.

\vskip 2 cm
\centerline{Figure Caption}
\vskip 1 cm
\noindent{Fig. 1. Transformation of the integration contour in the 
complex $\varphi$ plane.}

\begin{figure}
\leavevmode
\epsfxsize = 8.0truecm
\epsfysize = 7.2truecm
\includegraphics{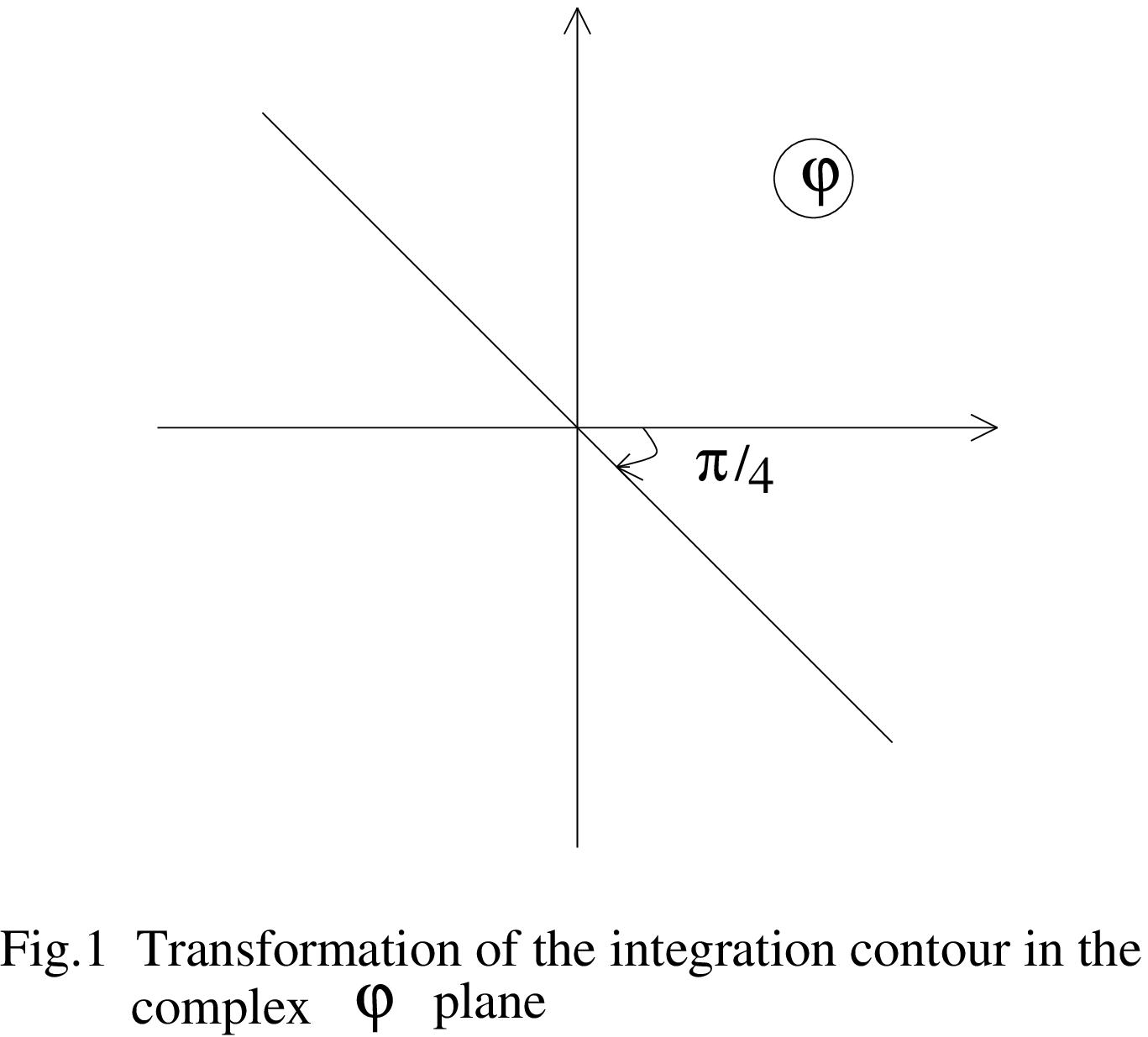}
\end{figure}

\end{document}